\documentclass[prd,nofootinbib,superscriptaddress,showpacs]{revtex4}

\usepackage{graphicx}
\usepackage{dcolumn}
\usepackage{amsmath}
\usepackage{epsfig}
\usepackage{amsfonts}
\usepackage{amssymb}
\usepackage{revsymb}

\begin{document}

\title{Constraints from the Detection of Cosmic Topology on the Generalized Chaplygin Gas}

\author{M. Makler}
\email{martin@cbpf.br} \affiliation{Observat\'orio Nacional - MCT, \\
Rua Gal. Jos\'e Cristino, 77 \\
20921-400, Rio de Janeiro -- RJ, Brazil}
\affiliation{Centro Brasileiro de Pesquisas F\'{\i}sicas \\
Rua Dr.\ Xavier Sigaud, 150 \\ 22290-180 Rio de Janeiro -- RJ,
Brazil}

\author{B. Mota}
\email{brunom@cbpf.br}

\author{M.J. Rebou\c{c}as}
\email{reboucas@cbpf.br}
\affiliation{Centro Brasileiro de Pesquisas F\'{\i}sicas \\
Rua Dr.\ Xavier Sigaud, 150 \\ 22290-180 Rio de Janeiro -- RJ, Brazil}

\date{\today}

\begin{abstract}
Despite our present-day inability to predict the topology of the
universe one may expect that we should be able to detect
it in the near future, given the increasing accuracy in the
astro-cosmological observations. Motivated by this, we examine to
what extent a possible detection of a non-trivial topology of a low
curvature ($\Omega_0 \sim 1$) universe, suggested by a diverse set
of current observations, may be used to place constraints on the
matter content of the universe, focusing our attention on the
generalized Chaplgygin gas (GCG) model, which unifies dark matter
and dark energy in a single matter component. We show that besides
constraining the GCG free parameters, the detection of a nontrivial
topology also allows to set bounds on the total density parameter
$\Omega_0$. We also study the combination of the bounds from the
topology detection with the limits that arise from  current data on
194 SNIa, and show that the determination of a given nontrivial
topology sets complementary bounds on the GCG parameters (and on
$\Omega_0$) to those obtained from the SNIa data.
\end{abstract}

\pacs{98.80.Es, 98.80.-k, 98.80.Jk}
%

\maketitle

\section{Introduction}

The isotropic expansion of the universe, the primordial abundance of
light elements, and the nearly uniform cosmic microwave background
radiation (CMBR) constitute the main observational pillars of the
standard Friedmann--Lema\^{\i}tre--Robertson--Walker (FLRW) model,
which provides a very successful description of the universe. In
this FLRW cosmological context, recent observations of high redshift
type Ia supernovae (SNIa) suggest that the universe is undergoing an
accelerated expansion. This picture is further strengthened by the
combination of recent CMBR observations (which imply that the total
density $\Omega_0$ is close or equal to unity) and the value
$\Omega_{m0} \simeq 1/3$ for the density of clustered (baryonic plus
dark) matter obtained from the $x$-ray emission in rich clusters of
galaxies, and from galaxy redshift surveys. This diverse set of
observations has led to an evolving consensus among cosmologists
that the universe is smoothly
permeated by a negative-pressure dark energy (DE) component~%
\cite{DEreviews}, which dominates the matter-energy of the universe
today ($\Omega_{DE} \simeq 2 \Omega_{m0} \simeq 2/3$), although it
must have been negligible in the past so as to permit structure
formation.

The nature of both dark matter (DM) and DE is still object of
intense investigations today. There are some candidates for DM from
particle physics, but yet no evidence of these suggested particles
has been found in laboratory experiments~\cite{DMreviews}. Regarding
DE there seems to exist no natural candidate from particle physics.
Thus, the current observational information regarding DM and DE
arises only from astro-cosmological observations. In addition to the
cosmological constant and a dynamical scalar field (quintessence;
see, e.g., refs.~\cite{Denergy}), the current paradigms for DE
include a number of possibilities (see, for example,
ref.~\cite{Alametal}),
among which the so-called generalized Chaplygin gas (GCG)%
~\cite{Kamenshchik,Makler,FabrisGCG,Bilic,BentoGCG,GCG1}, which
unifies DM and DE in a single matter-energy component, acting as cold
dark matter (CDM) at high redshifts and driving the accelerated
expansion today.
The behavior of the GCG in the framework of DM and DE
unification was extensively discussed in the literature (see, for
instance, refs.~\cite{Makler,FabrisGCG,Bilic,BentoGCG,GCG1,Sandvik,%
CMBGCG,GCG2,GCG4ess,skewness,reis03a,reis04b,FabrisSNIa,BertolamiSNIa,%
ZhuGCG,CMBRpheno,Dev4ess} and references therein).%
\footnote{The GCG model is in agreement with a number of
observational data related to the background dynamics, such as SNIa,
gravitational lenses, and Fanaroff-Riley Type IIb  galaxies (see
e.g. refs.~\cite{ZhuGCG,Dev4ess,GCG2}), but it fails to reproduce
large-scale structure data for adiabatic
perturbations~\cite{Sandvik}. However, for a specific type of
entropy perturbations the density fluctuation spectrum is consistent
with observational data~\cite{reis03a}. Note that these intrinsic
entropy perturbations are not ruled out by other current data (see
discussion in ref.~\cite{reis04b,skewness}). Thus, the (non
adiabatic) GCG is a viable framework to model the matter-energy
content of the universe.}

Somewhat parallel to these developments, and owing to the fact that
general relativity as a local metrical theory leaves undetermined
the space-time topology, a great deal of work  has also recently
gone into studying the possibility that the universe may possess
compact spatial sections with a non-trivial topology. On the one
hand,  methods and indicators have been devised to search for
topological signs of a possible non-trivial topology of our
$3$-space $M$ (see, e.g., refs.~\cite{TopSign,circles98}, and also
the review articles~\cite{CosmTopReviews} and~\cite{RG2004}). On the
other hand, the reported non-Gaussianity in CMBR
maps~\cite{CMB+NonGauss}, the small power of large-angle
fluctuations~\cite{WMAP-Spergel-et-al}, and some features in the
power spectrum~\cite{CMB+NonGauss,WMAP-Spergel-et-al} are
large-scale anomalies which have been suggested as a potential
indication of a non-trivial topology of the universe.  In this
regard the Poincar\'e dodecahedron space~\cite{Poincare} has been
suggested as an explanation for the weak wide-angle temperature
correlation in high precision CMBR data~\cite{WMAP}. However,
preliminary results~\cite{Cornishetal03}, using Wilkinson Microwave
Anisotropy Probe (WMAP) data, failed to find the six pairs of
matched circles of angular radius of about $35^{\circ}$ predicted by
the Poincar\'{e} space model~\cite{Poincare}. In this regard it is
important to note the results of the recent articles by Roukema
\emph{et al.}~\cite{Roukema} and Aurich
\emph{et al.}~\cite{Aurich1,Aurich2}, and some points made by Luminet%
~\cite{Luminet05}.

The immediate observational consequence of a multiply connected
$3$-space section $M$ of the universe is the existence of multiple
images of radiating sources: discrete cosmic objects or CMBR from
the last scattering surface. However, for these repetitions to be
detected the observable horizon radius $\chi_{hor}$ has to exceed at
least the smallest characteristic size $r_{inj}$ of the $3$-space
$M$. In this way, the question of detectability of cosmic (space)
topology naturally arises. This question has been recently studied
in the light of current astro-cosmological observations, which
indicate that our universe is nearly flat ($\Omega_0 \sim 1$; see,
e.g. ref.~\cite{SDSSWMAP}), and the extent to which a number of
non-trivial topologies may or may not be detected for the current
bounds on the cosmological density parameters has  been
determined in a few articles~\cite{Topdetect,Mota03,NEWweeks,%
NEWweeks2}.

These studies have concentrated in the $\Lambda$CDM
framework, where the cold dark (and baryonic) matter plus a
cosmological constant ($\Lambda$) are the matter-energy
constituents.
In a recent article~\cite{MMR1} the detectability of cosmic topology
of low curvature universes ($\Omega_0 \sim 1$) has been
discussed in the unified DM and DE GCG context.

The main aim of this paper is to address the detectability of cosmic
topology inverse problem, i.e. to investigate the extent to which
the detection of a non-trivial nearly flat ($\Omega_0 \sim 1$)
topology may constrain models for the matter-energy content of the
universe. To this end, we shall focus our attention on the GCG model
and derive bounds set on its two free parameters from the detection
of several possible compact topologies. We also consider the
combination of the bounds from the topology detection  with the
limits on the GCG parameters which arise from current data from 194
SNIa. It is found that detection of a nontrivial topology (through
CMBR pattern repetitions) sets complementary limits to those
obtained from SNIa data. In particular, the detection of some
specific manifolds taken together with current SNIa data may place
constraints comparable to those expected from space based
experiments (such as the Supernovae Acceleration Probe).

We also show that besides limiting the GCG parameters, the detection
of a nontrivial topology allows to set bounds on the total density
parameter $\Omega_0$, which are shown to be further constrained by the
combination of topology detection with the current SNIa data.
Such bounds on  $\Omega_0$ can be confronted with the values which
arise from some topology detection methods (such as the circles
in-the-sky~\cite{circles98}) in order to have further
limits on the GCG parameters.

The outline of the paper is as follows. In the next section, we give
a brief account of the detectability of cosmic topology basic
context and review a few topological results regarding the
$3$-manifolds, which will be used in the following sections.
Focusing on the GCG as a matter content model, we discuss in
section~\ref{DetectInver} the cosmic topology inverse problem and
set bounds on the GCG parameters from the possible detection of
several compact topologies. The combination of cosmic topology
detection with current SNIa data and its implications for
constraining the GCG parameters and $\Omega_0$ are discussed in
section~\ref{combSNIa}. Finally, in section~\ref{Conclusion} we
discuss our main results and present some concluding remarks. A few
details of the supernovae analysis are discussed in the appendix.

\section{Detectability Problem in Cosmic Topology}
\label{detect}
To make the article as clear and self-contained as possible,
in this section we shall present the cosmic topology basic context,
state the detectability condition, and recall some topological
properties of $3$-manifolds which will be used in the following
sections.

Within the framework of the standard FLRW cosmology, the universe is
modeled by a $4$-manifold $\mathcal{M}$ which is decomposed into
$\mathcal{M} = \mathbb{R} \times M$, and is endowed with a locally
isotropic and homogeneous Robertson--Walker (RW) metric
\begin{equation}
\label{RWmetric} ds^2 = -dt^2 + a^2 (t) \left [ d \chi^2 +
f^2(\chi) (d\theta^2 + \sin^2 \theta  d\phi^2) \right ] \;,
\end{equation}
where $f(\chi)=(\chi\,$, $\sin\chi$, or $\sinh\chi)$ depending on
the sign of the constant spatial curvature ($k=0,1,-1$), and $a(t)$
is the scale factor.

The spatial section $M$ is usually taken to be one of the following
simply-connected spaces: Euclidean $\mathbb{E}^{3}$ ($k=0$),
spherical $\mathbb{S}^{3}$ ($k=1$), or hyperbolic $\mathbb{H}^{3}$
($k=-1$) spaces. However, since geometry does not dictate topology,
the $3$-space $M$ may equally well be any one of the possible
quotient (multiply connected) manifolds $M = \widetilde{M}/\Gamma$,
where $\Gamma$ is a discrete and fixed point-free group of
isometries of the covering space
$\widetilde{M}=(\mathbb{E}^{3},\mathbb{S}^{3}, \mathbb{H}^{3})$. In
forming the quotient manifolds $M$ the essential point is that they
are obtained from $\widetilde{M}$ by identifying points which are
equivalent under the action of the discrete group $\Gamma$. Hence,
each point on the quotient manifold $M$ represents all the
equivalent points on the covering manifold $\widetilde{M}$. The
action of $\Gamma$ tessellates (tiles) $\widetilde{M}$ into
identical cells or domains which are copies of what is known as
fundamental polyhedron (FP).%
\footnote{A simple example of quotient manifold in two dimensions is
the $2$--torus $T^2 = \mathbb{S}^1 \times \mathbb{S}^1=
\mathbb{E}^2/ \Gamma$. The  covering space clearly is
$\mathbb{E}^2$, and a FP is a rectangle with opposite sides
identified. This FP tiles the covering space $\mathbb{E}^2$. The
group $\Gamma=\mathbb{Z}\times \mathbb{Z}$ consists of discrete
translations associated with the side identifications.}

In a multiply connected manifold, any two given points may be joined
by more than one geodesic. Since the radiation emitted by cosmic
sources follows (space-time) geodesics, the immediate observational
consequence of a non-trivial spatial topology of $M$ is that the sky
may (potentially) show multiple images of radiating sources: cosmic
objects or specific correlated spots of the CMBR. At large
cosmological scales, the existence of these multiple images (or
pattern repetitions) is a physical effect often used to examine the
detectability of the $3$-space topology.

In order to state the conditions for the detectability of cosmic
topology in the context of standard cosmology, we note that for
non-flat metrics of the form~(\ref{RWmetric}), the scale factor
$a(t)$ can be identified with the curvature radius of the spatial
section of the universe at time $t$. Therefore $\chi$ is the
distance of any point with coordinates $(\chi, \theta, \phi)$ to the
origin (in the covering space) \emph{in units of the curvature
radius}, which is a natural unit of length that shall be used
throughout this paper.

The study of the detectability of a possible non-trivial topology of
the spatial section $M$ requires a topological typical length which
can be put into correspondence with observation survey depths
$\chi_{obs}$ up to a redshift $z=z_{obs}$. A suitable characteristic
size of $M$, which we shall use in this paper, is the so-called
injectivity radius $r_{inj}$, which is nothing but the radius of the
smallest sphere `inscribable' in $M$, and is defined in terms of the
length of the smallest closed geodesics $\ell_M\,$ by $r_{inj} =
\ell_M/2$.

Now, for a given survey depth $\chi_{obs}$ a topology is said to be
undetectable if $\chi_{obs} < r_{inj}$. In this case there are no
multiple images (or pattern repetitions of CMBR spots) in the survey
of depth $\chi_{obs}$. On the other hand, when $\chi_{obs} >
r_{inj}$, then the topology is detectable in principle or
potentially detectable.

In a globally homogeneous manifold the above detectability condition
holds regardless of the observer's position, and so if the topology
is potentially detectable (or is undetectable) by an observer at $x
\in M$, it is potentially detectable (or is undetectable) by an
observer at any other point in the $3$-space $M$. However, in
globally inhomogeneous manifolds the detectability of cosmic
topology depends on both the observer's position $x$ and the survey
depth $\chi_{obs}$. Nevertheless, even for globally inhomogeneous
manifolds the above defined `global' injectivity radius $r_{inj}$
can be used to state an \emph{absolute\/} undetectability condition,
namely $r_{inj} > \chi_{obs}$. Reciprocally, the condition
$\chi_{obs} > r_{inj}$ allows potential detectability (or
detectability in principle) in the sense that, if this condition
holds, multiple images of topological origin are potentially
observable at least for some observers suitably located in $M$. An
important point is that for spherical and hyperbolic manifolds,
which we focus on in this work, the `global' injectivity radius
$r_{inj}$ expressed in terms of the curvature radius, is a constant
(topological invariant) for a given topology.

In the remainder of this section we shall recall some relevant
results about spherical and hyperbolic $3$-manifolds, which will be
useful in the following sections. The multiply connected spherical
$3$-manifolds are of the form $M=\mathbb{S}^3/\Gamma$, where
$\Gamma$ is a finite subgroup of $SO(4)$ acting freely on the
$3$-sphere. These manifolds were originally classified  by Threlfall
and Seifert~\cite{ThrelfallSeifert}, and are also discussed by
Wolf~\cite{Wolf} (for a description in the context of cosmic
topology see~\cite{Ellis71}). Such a classification consists
essentially in the enumeration of all finite groups $\Gamma \subset
SO(4)$, and then in grouping the possible manifolds in classes. In a
recent paper, Gausmann \emph{et al.\/}~\cite{GLLUW} recast the
classification in terms of single action, double action, and linked
action manifolds. In table~\ref{SingleAction} we list the single
action manifolds together with the symbol often used to refer to
them, as well as the order of the covering group $\Gamma$ and the
corresponding injectivity radius. It is known that single action
manifolds are globally homogeneous, and thus the detectability
conditions for an observer at an arbitrary point $p \in M$ also hold
for an observer at any other point $q \in M$. Finally we note that
the binary icosahedral group $I^{\ast}$ gives the known Poincar\'e
dodecahedral space, whose fundamental polyhedron is a regular
spherical dodecahedron, $120$ of which tile the $3$-sphere into
identical cells which are copies of the FP.

\begin{table}[!htb]
\begin{center}
\begin{tabular}{|l|c|c|c|}
\hline
\ Name \&  Symbol \ &  \ Order of $\Gamma$ \ & \ Injectivity Radius \  \\
\hline \hline
Cyclic              $Z_n$   & $n$  & $\pi/n$           \\
Binary dihedral     $D^*_m$ & $4m$ & $\pi / 2m $      \\
Binary tetrahedral  $T^*$   & 24   & $ \pi/6$          \\
Binary octahedral   $O^*$   & 48   & $\pi/8$          \\
Binary icosahedral  $I^*$   & 120  & $\pi/10$         \\
\hline
\end{tabular}
\end{center}
\caption{Single action spherical manifolds together with the order
of the covering group and the injectivity radius.}
\label{SingleAction}
\end{table}

Despite the enormous advances made in the last few decades, there is
at present no complete classification of hyperbolic $3$-manifolds.
However, a number of important results have been obtained, including
the two important theorems of Mostow~\cite{Mostow} and
Thurston~\cite{Thurston}. According to the former, geometrical
quantities of orientable hyperbolic manifolds, such as their volumes
and the lengths of their closed geodesics, are topological
invariants. Therefore quantities such as the `global' injectivity
radius $r_{inj}$ (expressed in units of the curvature radius) are
fixed for each manifold. Clearly this property also holds for
spherical manifolds. 

According to Thurston's theorem, there is a countable infinity of
sequences of compact orientable hyperbolic manifolds, with the
manifolds of each sequence being ordered in terms of their volumes.
Moreover, each sequence has as an accumulation point a cusped
manifold, which has finite volume, is non-compact, and has
infinitely long cusped corners~\cite{Thurston}.
%
\begin{table}[!htb]
\begin{center}
\begin{tabular}{|l|c|c|c|}
\hline
\ \ \ Manifold \ \ \ & \ \ \ Volume \ \ \ & \ \ \ Injectivity Radius  \\
\hline \hline
\ \ m004(1,2) &  1.398  & 0.183  \\ 
\ \ m004(6,1) &  1.284  & 0.240  \\ 
\ \ m003(-4,3) &  1.264  & 0.287  \\ 
\ \ m003(-2,3) &  0.981  & 0.289  \\ 
\ \ m003(-3,1) &  0.943  & 0.292  \\ 
\ \ m009(4,1) &  1.414  & 0.397  \\ 
\ \ m007(3,1) &  1.015  & 0.416  \\ \hline
\end{tabular}
\caption[Hodgson-Weeks census.] {\label{10HW-Census}
First seven manifolds in the Hodgson-Weeks
census of closed hyperbolic manifolds, ordered by the
injectivity radius $r_{inj}$, together with their
corresponding volume.}
\end{center}
\end{table}

Closed orientable hyperbolic $3$-manifolds can be constructed from
these cusped manifolds. The compact manifolds are obtained through a
so-called Dehn surgery which is a formal procedure identified by two
coprime integers, i.e. winding numbers $(n_1,n_2)$. These manifolds
can be constructed and
studied with the publicly available software package SnapPea%
~\cite{SnapPea}. SnapPea names manifolds according to the seed
cusped manifold and the winding numbers. So, for example, the
smallest volume hyperbolic manifold known to date (Weeks' manifold)
is named as m$003(-3,1)$, where m003 corresponds to a seed cusped
manifold, and $(-3,1)$ is a pair of winding numbers. Hodgson and
Weeks~\cite{SnapPea,HodgsonWeeks} have compiled a census containing
$11031$ orientable closed hyperbolic 3-manifolds ordered by
increasing volumes. In table~\ref{10HW-Census} we collect the first
seven manifolds from this census with the lowest volumes, ordered by
increasing injectivity radius $r_{inj}$, together with their
volumes.

\section{Detectability Inverse Problem}
\label{DetectInver}

The detectability of cosmic topology as well as its inverse
problem can be said to have two main ingredients, namely one of
mixed geometric-and-topological nature, and another which comes
from the fact the current astro-cosmological observations allow a
set of possible models for the matter-energy content of the
universe. As we have discussed in the previous section, the former
arises from the fact that RW geometry does not dictate the
topology of the $3$-space $M$, giving rise to a multiplicity of
non-trivial topologies for $M$ and the possible existence of
multiple images of radiating sources.

Regarding the second important ingredient, we shall focus on the GCG
unification of DM and DE paradigm to concretely illustrate the
cosmic topology inverse problem. In other words, we assume the
current matter content of the universe to be given by the ordinary
baryonic matter of density $\rho_b$ plus a generalized Chaplygin gas
(with density $\rho_{ch}$ and pressure $p_{ch}$) whose equation of
state is given by~\cite{FabrisGCG,Kamenshchik,Makler,Bilic,BentoGCG}

\begin{equation}
\label{StateEq}
p_{\,ch} = - \frac{M^{4(\alpha+1)}}{\rho^\alpha_{ch}} \;,
\end{equation}
where the constant $M>0$ has dimension of mass and $\alpha$
is a real dimensionless constant. The Friedmann equation is
then given by%
\begin{equation}
\label{Friedmann}
H^2 =\frac{8 \pi G}{3}\,\,(\rho_b+\rho_{ch})-\frac{k}{a^2} \;,
\end{equation}
where $H=\dot{a}/a$ is the Hubble parameter, overdot stands for
derivative with respect to time $t$, and $G$ is Newton's
constant. Introducing the critical density
$\rho_{\mathrm{cr}} = 3 H^2\! / (8\pi G)\,$, and the corresponding
densities $\Omega_b = \rho_b / \rho_{\mathrm{cr}}\,$ and
$\Omega_{ch} = \rho_{ch} / \rho_{\mathrm{cr}}\,$, equation%
~(\ref{Friedmann}) can be rewritten
as
\begin{equation}
\label{Friedmann2}
a^2 H^2 (\Omega - 1) = k  \;,
\end{equation}
where $\Omega = \Omega_b + \Omega_{ch}\,$.

If one further assumes that these two matter components do
not interact, then the energy conservation equation
$\dot{\rho}_{\,tot} + 3\,H (\rho_{\,tot} + p_{\,tot})=0\,$ can be
integrated separately for the baryonic matter and Chaplygin gas,
giving the well known result $\rho_b = \rho_{b0} (a_0/ a)^3\,$,
and
\begin{equation}
\label{rhoch}
\rho_{ch} = \rho_{ch0} \left[(1-A)\left(\frac{a_{0}}{a}\right)^{3(1+\alpha)}
+A\right]^{1/(1+\alpha)},
\end{equation}
where $A=(M^{4} /\rho_{ch0})^{(1+\alpha)}$, and the index $0$
denotes evaluation at present time $t_0$. We note that at high
redshifts ($a_0/a \gg 1$) one has $\rho_{ch} \propto a^{-3}$,
whereas at late times ($a_0/a \ll 1$) one has $p_{ch} = -\rho_{ch} =
-M^{4} = const.$, making clear that the GCG interpolates between a
dust dominated (CDM) phase in the past, and a cosmological constant
phase in the future.

Writing the above expressions for the densities $\rho_b$ and
$\rho_{ch}$ in terms of the redshift ($z=a_0/a-1$), the Friedmann
equation~(\ref{Friedmann2}) can be rearranged to give the Hubble
function
\begin{equation}
\label{HubbleFun}
H(z) = H_0 \left\{  \Omega_{ch0} \left[ A +(1-A)(1+z)^{3(1+\alpha)}
\right]^{1/(1+\alpha)}  + \Omega_{b0} (1+z)^3 +(1-\Omega_0)(1+z)^2
           \right\}^{1/2} \;,
\end{equation}
where $\Omega_{0} = \Omega_{ch0}+ \Omega_{b0}$. It is clear from
this equation that, regardless of the value of $\alpha$, for $A=0$
the GCG component behaves as CDM with density $\Omega_{ch0}$, while
for $A=1$ the GCG plays the role of a cosmological constant term,
whose density is again $\Omega_{ch0}$. On the other hand, for $\alpha=0$
the GCG behaves as $\Lambda$CDM, whose matter and cosmological constant
components are, respectively, given by $\Omega_{ch0} (1-A)$ and
$\Omega_{ch0} A$.

From (\ref{HubbleFun}) we find that the redshift-distance relation
for non-flat cases, in units of curvature radius $a_0$, is given by
\begin{equation}
\label{ChiObsGCG1}
\chi =
\sqrt{|1-\,\Omega_{0}|}\,
\int_{1}^{1+z}\!\left\{
\Omega_{ch0}\left[\,A+(1-A)\,x^{3(1+\alpha)}\right]^{1/(1+\alpha)}
+\Omega_{b0}\,x^{3} + (1-\Omega_{0})\,x^{2}
               \right\}^{-1/2} dx   \;,
\end{equation}
where $x=1+z$ is an integration variable and we have used that, for
non-flat models ($k \neq 0$), the curvature radius is identified with
the scale factor, which from~(\ref{Friedmann2}) is
given by $a_0= 
(\,H_0\sqrt{|1-\Omega_0|}\,)^{-1}\,$  today. 
For simplicity, on the left hand side of~(\ref{ChiObsGCG1})
and in many places in the remainder of this article, we have
left implicit the dependence of the function $\chi$ on
its variables.

For a given survey with redshift cut-off $z_{obs}$, the redshift
distance function $\chi_{obs}$ clearly depends on the way one
models the matter-energy content of the universe. So, for example,
in the $\Lambda$CDM context $\chi_{obs} =
\chi\,(\Omega_{m0},\Omega_{\Lambda0})$, and therefore the
potential detectability (or the undetectability) of a given
topology depends on these density parameters.
Similarly, in dealing with the detectability of cosmic topology in
the GCG unified framework it is clear from~(\ref{ChiObsGCG1}) that
besides the Chaplygin density $\Omega_{ch0}$, one has to consider
the GCG model parameters $\alpha$ and $A$, assuming we know
$\Omega_{b0}$. Unless otherwise stated, we shall assume in what
follows that $\Omega_{b0}=0.04$, which is the value that arises
from recent observations.%
\footnote{The determination of light element abundances together
with primordial Big-Bang nucleosynthesis~\cite{BBN,DH} furnishes
$\Omega _{b0} h^{2}=0.0214 \pm 0.0018$, which combined with $h=0.72
\pm 0.08$ ($H_{0} \equiv 100\,h\,Km/s/Mpc$), from the
Hubble Space Telescope Key Project~\cite{Freedman}, gives
the central value $\Omega_{b0}\simeq0.04$. Although these results
were derived in the $\Lambda$CDM context, the values remain
unaltered in the GCG unified framework, since the GCG behaves as CDM
at the relevant (high) redshifts and the observational determination
of the $H_{0}$ (made through the linear Hubble law) is also
unaffected by the nature of the DM and DE
components.}

To examine how a possible detection of cosmic topology may
constrain the GCG parameters, consider a  manifold $M$ with
$r_{inj}=r_{inj}^{\:M}\,$ (say), and a given survey depth
$z=z_{obs}$. It is clear that, for a fixed value of a variable of
the redshift-distance function~(\ref{ChiObsGCG1}), equation
\begin{equation}
\label{ContCurves}
\chi_{obs} (A,\alpha, \Omega_0) = r_{inj}^{\:M}  
\end{equation}
defines an implicit function of the remaining two parameters, whose
graph is a curve in the corresponding parameter plane. In the
parameter planes $\Omega_0$--$A$ and $\,\Omega_0$--$\,\alpha$ for
all points above (below) a curve solution of~(\ref{ContCurves}), the
topology of the corresponding spherical (or respectively hyperbolic)
manifold is potentially detectable.%
\footnote{In the $\,A$--$\,\alpha$ plane, for points above the curves
solution of (\ref{ContCurves}), the topology is detectable for both
spherical and hyperbolic manifolds.}
Obviously, for all points below (or above in the hyperbolic case)
these curves, the corresponding spherical (hyperbolic) topologies
are undetectable ($\chi_{obs} < r_{inj}$). In this way, the
detection of a non-trivial cosmic topology $M$ with injectivity
radius $r_{inj}^{\:M}$ may be used to set constraints on the GCG
parameters $\alpha$ and $A$, for $\Omega_0$ given by recent
observations.

Before we proceed further we note that, for $\rho_{ch}$ to be well
defined for any value of the scale factor, the parameter $A$ has to
be limited to the interval $0 \leq A \leq 1$. Regarding $\alpha$, we
must have $\alpha > -1 $, so as to ensure that the GCG behaves as CDM
at early times and as a positive cosmological constant at late times.
As we shall see in section~\ref{combSNIa}, recent SNIa data give
$\alpha\lesssim 12$ (see also refs.~\cite{BertolamiSNIa,CMBRpheno}).
Thus, in what follows, we shall take $-1<\alpha <12$.

As a first concrete example of bounds set by the detection of cosmic
topology on the GCG model parameters, we consider the behavior of
these curves solution for fixed values of $\alpha$ for some single
action spherical and small volume hyperbolic manifolds of
Tables~\ref{SingleAction}
and~\ref{10HW-Census},%
\footnote{As the three manifolds in table~\ref{10HW-Census} obtained
from the cusped seed manifold m$003$ have very close injectivity
radii, for the sake of simplicity we shall present only the results
corresponding to Weeks manifold m$003(-3,1)$ in the remainder of
this work.}
taking into account the redshift $z=1089$ of the last scattering
surface \cite{WMAP-Spergel-et-al}. As an illustration, we shall
consider the interval $0.99< \Omega_0 <1.03$, which arises from a
combined analysis of CMBR, SNIa, and large-scale structure
data~\cite{SDSSWMAP} (at 68\% confidence level) in the $\Lambda$CDM
framework.

\begin{figure}[!hbt]
\centerline{\def\epsfsize#1#2{1.1#1}\epsffile{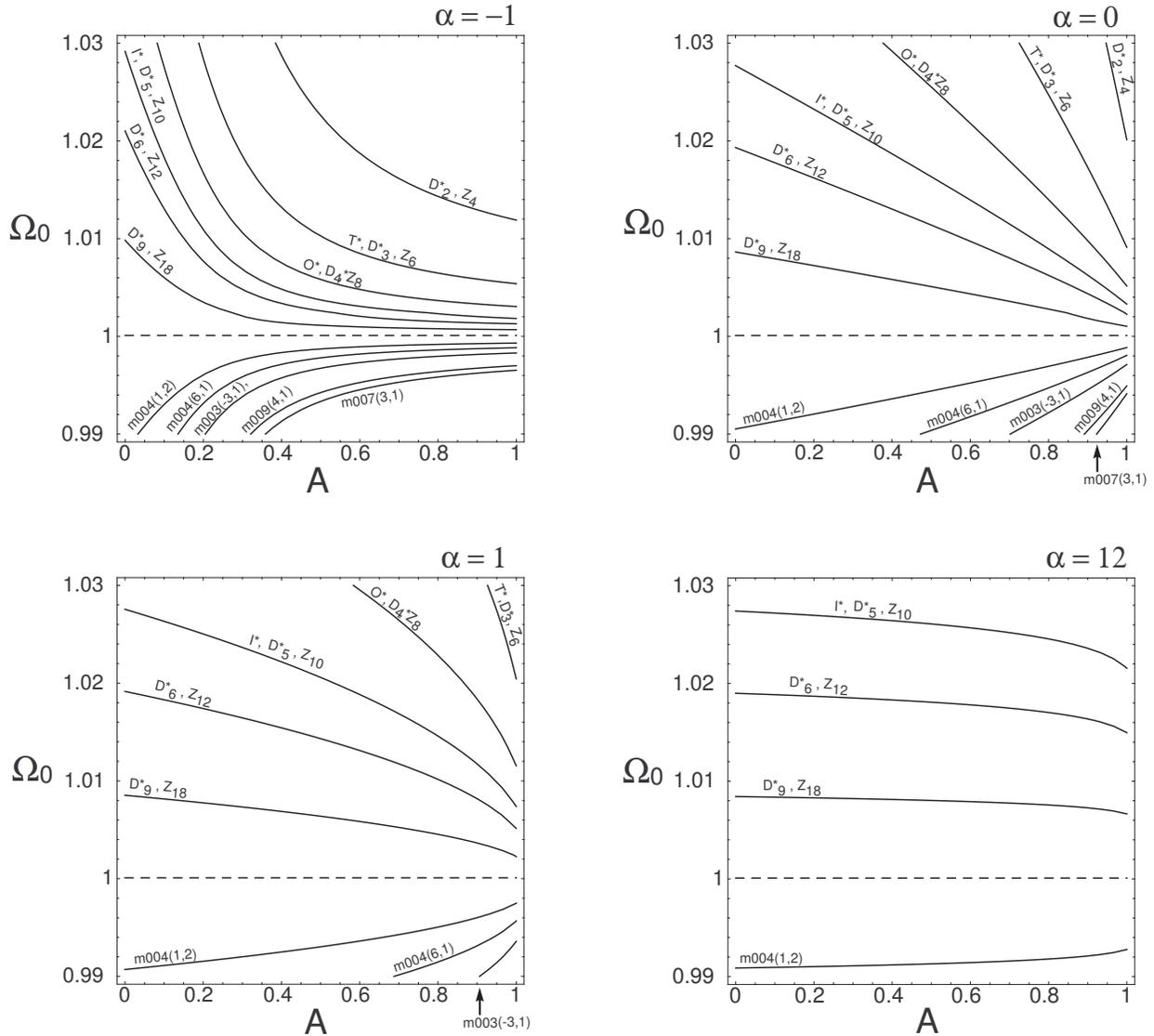}}
\caption{\label{OmegaVersusA} The solution curves
of~(\ref{ContCurves}) as plots of $\Omega_0$ versus $A$ for fixed
values of $\alpha$ and $r_{inj}$ of some spherical and hyperbolic
manifolds of tables~\ref{SingleAction} and~\ref{10HW-Census}. A
survey depth $z_{obs}= 1089$ (CMBR) was used in all cases.}
\end{figure}

Figure~\ref{OmegaVersusA} gives plots of the curves solution of
$\chi_{obs} (\Omega_0,A) = r_{inj}$, for $\alpha = -1, 0, 1$, and
$12$ and $\Omega_{b0}=0.04$. Given that each of these curves
separates the parameter plane into (potentially) detectable and
undetectable regions, it is clear that, for example, the detection
of a binary tetrahedral $T^*$ topology (or binary dihedral $D^*_3$,
or cyclic $Z_6$, which have the same $r_{inj}$) would set the bounds
$A \gtrsim 0.2$, $A \gtrsim 0.7$, and $A \gtrsim 0.9$, for
$\alpha=-1$, $\alpha=0$ ($\Lambda$CDM), and $\alpha=1$,
respectively. On the other hand, the detection of the Poincar\'e
dodecahedron space topology would not set any constraint on $A$ for
the considered upper bound on $\Omega_0$.

We note that, according to equation~(\ref{ChiObsGCG1}), the smaller
the value of $|\Omega_0 - 1|$ the greater is the number of
undetectable topologies. However, for any value of $|\Omega_0 - 1|$,
no matter how small, there will always remain an infinite number of
topologies in the binary dihedral $D^*_m$ and cyclic $Z_n$ families,
which are in principle detectable for large order covering groups
$\Gamma$ (see Table~\ref{SingleAction}). Nevertheless,
figure~\ref{OmegaVersusA} makes clear that for $n
> 10$ and $m > 5$ the detection of topology (using CMBR) puts no
constraints on $A$, for the values of $\alpha$ consistent with
recent SNIa observations, with $\Omega_0$ within the interval given
by CMBR and large-scale structure data.

\begin{figure}[!hbt]
\centerline{\def\epsfsize#1#2{0.9#1}\epsffile{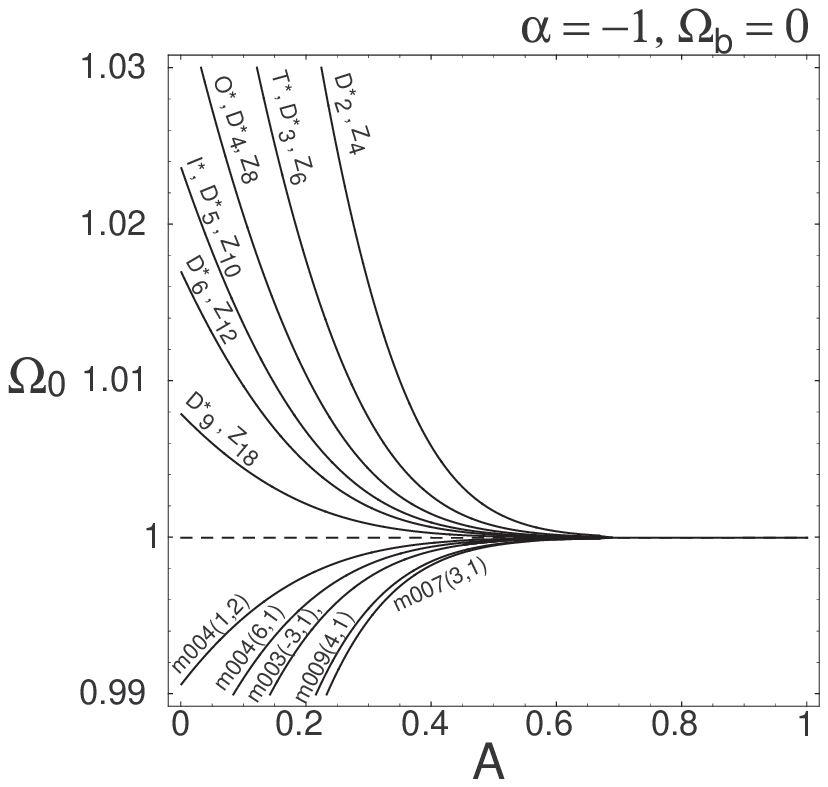}}
\caption{\label{OmegaVersusAbarion} The solution curves of
$\chi_{obs}=r_{inj}$ for $\alpha=-1$ and  $r_{inj}$ of
some spherical and  hyperbolic manifolds of tables~\ref{SingleAction}
and~\ref{10HW-Census}. A survey depth $z_{obs}= 1089$ (CMBR) was used.
This figure illustrates the role played  by the baryonic matter in the
detectability of cosmic topology and its inverse problem.}
\end{figure}

The picture is rather similar for hyperbolic manifolds. For example,
the detection of the Week's manifold m$003(-3,1)$ would give the
bounds $A \gtrsim 0.2$, $A \gtrsim 0.7$ and $A \gtrsim 0.9$ for,
respectively, $\alpha=-1$, $\alpha=0$, and $\alpha=1$. The smaller
the injectivity radius the weaker is the restriction on $A$ that
arises from the detection of the topology, for any fixed $\alpha$.
So, for example, for $\alpha =1$,  the detection of the m$004(1,2)$
topology imposes no bounds on $A$, while for the manifold
m$004(6,1)$ it gives $A \gtrsim 0.7$.

Finally, from figure~\ref{OmegaVersusA} one also has that in general
the greater is $\alpha$ the smaller is the interval allowed for $A$
as consequence of a possible detection of spherical and hyperbolic
topologies.

An important point to be noticed here is that since $\chi_{obs}$ is
independent of $\alpha$ for $A=1$, according to~(\ref{ChiObsGCG1}),
there is an absolute minimum (maximum) of $\Omega_0$ which arises
from the detection of a given spherical (hyperbolic) manifold.
This clearly illustrates how bounds on a local physical quantity
can be imposed by the global topology of the universe.
The fact that a possible detection of a nontrivial topology places
constraints on $\Omega_0$ was first mentioned in the cosmological
context by Bernshtein and Shvartsman~\cite{BernshteinShvartsman1980},
and has been recently dealt with in the $\Lambda$CDM framework by
Roukema and Luminet~\cite{RoukemaLuminet1999}. We shall show in the
next section that the combination of cosmic topology detection with
bounds from other experiments, such as SNIa data, allows to place
even stronger limits on $\Omega_0$.

The role played by the baryonic matter component is not significant
for positive values of $\alpha$ (and $A$ not too close to $1$), but
is very important for $-1 \leq \alpha < 0$, as
figure~\ref{OmegaVersusAbarion} demonstrates for $\alpha=-1$. The
comparison of this figure with figure~\ref{OmegaVersusA}a clearly
shows that the detection of any spherical or hyperbolic topology
would set less stringent bounds on $A$ in the absence of the
baryonic matter.

\begin{figure}[!hbt]
\centerline{\def\epsfsize#1#2{0.9#1}\epsffile{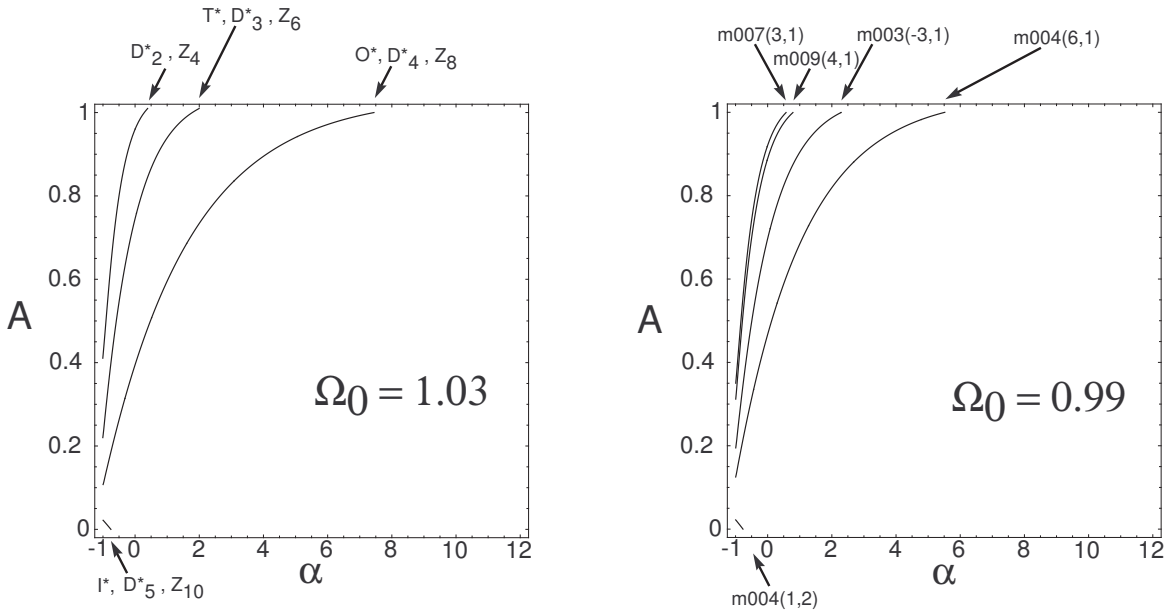}}
\caption{\label{AVersusAlfa} The solution curves of
$\chi_{obs}=r_{inj}$ for two limiting values of the total
density, 
and $r_{inj}$ of some spherical and hyperbolic manifolds
of tables~\ref{SingleAction} and~\ref{10HW-Census}. The redshift
survey depth of the last scattering surface ($z_{obs}= 1089$) was
used.}
\end{figure}

We shall now focus our attention on the parameter plane
$A$--$\alpha$. Figure~\ref{AVersusAlfa} gives plots of the curves
solution of $\chi_{obs} (A,\alpha) = r_{inj}$, for the limiting
values of the total density parameter $\Omega_0$ and
$\Omega_{b0}=0.04$. The left panel in the figure shows the curves
of some single action spherical manifolds
(table~\ref{SingleAction}), while on the right panel we have the
curves of some small hyperbolic manifolds
(table~\ref{10HW-Census}).

From the region above the curves of single action manifolds one
reads that, for example, the detection of a $D^*_3$ topology (or
equivalently $T^*$ and $Z_6$) gives the bounds $\alpha \lesssim 2$
and $A \gtrsim 0.2$ (for $\Omega_0 =1.03$). The Poincar\'e
dodecahedron (as well as $D^*_5$ and $Z_{10}$) is detectable for
most of the range of $A$ and $\alpha$. Its detection only places the
limits $A \gtrsim 0.05$ and $\alpha \gtrsim -0.5$. For the binary
dihedral $D^*_m$ and cyclic $Z_n$ families, figure~\ref{AVersusAlfa}
makes clear that the detection of topology puts no constraints on
both $\alpha$ and $A$, using CMBR ($z=1089$), for $n > 10$ and $m >
5$, for the limiting value $\Omega_0=1.03$.

For $\Omega_0 = 0.99$, the detection of the smallest (known)
hyperbolic manifold m$003(-3,1)$ would set the constraints $\alpha
\lesssim 2.5$ and $A \gtrsim 0.2$, while the detection of
m$004(1,2)$ gives almost no restriction on both GCG parameters.

Hitherto we have examined the detectability of cosmic topology
inverse problem in terms of the parameters $A$ and $\alpha$, which
are directly constrained by some experiments, such as SNIa
observations. However, these experiments probe the redshift-distance
function~(\ref{ChiObsGCG1}), and do not give rise to strong
constraints on that parameters using current data (see,
e.g.,~\cite{GCG2}). On the other hand, observables related to the
clustering of matter allow to place tighter limits on the GCG
parameters~\cite{GCG2}, and are sensitive to the part of the GCG
that clumps. In this way, to proceed further with our study of the
inverse problem we identify
the clumping matter density parameter with%
\footnote{Another possible choice of the matter density parameter, which
has been made in ref.~\cite{MMR1}, is $\bar{\Omega}_{m0}= \Omega_{ch0}\,(1-A)$.
In this case only for $\alpha=0$ this density scales as $a^{-3}$
for $a \ll 1$.}
\begin{equation} \label{NewVar}
\bar{\Omega}_{m0}= \Omega_{ch0}\,(1-A)^{1/(1+\alpha)}\;,
\end{equation}
such that in the Hubble function~(\ref{HubbleFun}) both the baryonic
matter and the GCG scale as $(1+z)^3$ at high redshifts. In this
way, the density $\bar{\Omega}_{m0}$ corresponds to the fraction of
the GCG that behaves as CDM, and therefore an effective mass density
parameter $\Omega_m^{eff}= \bar{\Omega}_{m0} + \Omega_{b0}$ can be
identified as the fraction of the total density that
evolves as pressureless matter.%
\footnote{The identification~(\ref{NewVar}) observationally well
motivated as the effective matter density of the GCG.
Indeed, using $\Omega_m^{eff}$ as the cold dark matter density
parameter in the initial fluctuation spectrum (and assuming entropy
perturbations) leads to a large-scale matter power spectrum in
agreement with observational data~\cite{reis03a}. Furthermore, the
values of $\Omega_m^{eff}$ obtained from $x$-ray gas mass fraction
in clusters~\cite{GCG2} are consistent with those derived from the
power spectrum.}
In terms of $\bar{\Omega}_{m0}$ the redshift-distance relation%
~(\ref{ChiObsGCG1}) takes the form
\begin{equation}
\label{ChiObsGCG2}  
\chi \left(\Omega_0,\bar{\Omega}_{m0},
    \alpha \right) =
\sqrt{|1-\Omega_{0}|} \, \int_{1}^{1+z}\!\! \left\{ \Omega_{ch0}
\left[ 1 -
\left(\frac{\bar{\Omega}_{m0}}{\Omega_{ch0}}\right)^{\!\!1+\alpha}
+\left(\frac{\bar{\Omega}_{m0}}{\Omega_{ch0}}
\,x^3\right)^{\!\!1+\alpha} \right]^\frac{1}{1+\alpha} \!\!\!\!+
\Omega_{b0}\,x^{3} + (1-\Omega_{0})\,x^{2} \right\}^{\!\!-1/2}
\!\!\!\!\!\!\!\! dx \;.
\end{equation}

\begin{figure}[!hbt]
\centerline{\def\epsfsize#1#2{1.0#1}\epsffile{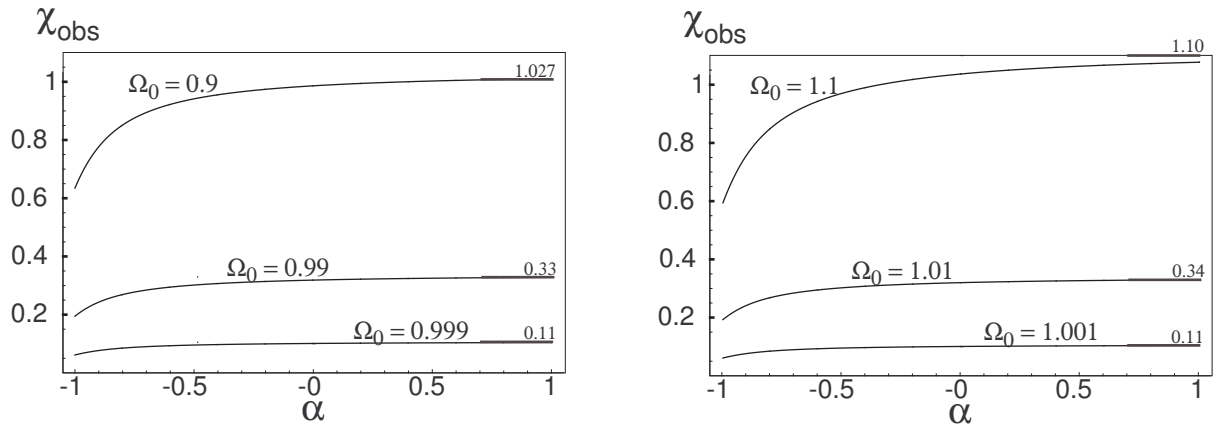}}
\caption{\label{ChiVersusAlfa} The behavior os $\chi_{obs}$ as
function of $\alpha$ for different values of total density
$\Omega_0$, for $\Omega_m^{eff} = 0.3$ and $\Omega_{b0} = 0.04$. The
redshift $z_{obs}= 1089$ of the last scattering surface was used.}
\end{figure}

Assuming that $\Omega_m^{eff}$ is fixed by some independent
measurement, say the $x$-ray gas mass fraction in clusters, for a
given survey depth $z_{obs}$ we have
$\chi_{obs}=\chi_{obs}(\Omega_0, \alpha)$, which can be shown to be
a monotonically increasing function of $\alpha$, for $\alpha
\in(-1,\infty)$, and any fixed total density.
Figure~\ref{ChiVersusAlfa} illustrates this behavior for different
values of $\Omega_0$, for $\Omega_m^{eff}=0.3$,
$\Omega_{b0}=0.04\,$, and for the redshift corresponding to the last
scattering surface ($z_{obs}=1089$). At first sight, the
monotonically increasing behavior of $\chi_{obs}$ indicates that the
detectability of a given topology becomes more likely as $\alpha$
increases. However, for $\alpha>1$ the behavior of $\chi_{obs}$ is
not very sensitive to $\alpha$ as can be inferred from the
asymptotic values of $\chi_{obs}$, whose values are indicated in the
figure by small line segments. On the other hand, $\chi_{obs}$
changes substantially for $\alpha \lesssim -0.5$, and for universes
that are not so nearly flat, i.e. for  $ 0.01 \lesssim |\Omega_0 -1|
\lesssim 0.1$. As a consequence, for $\alpha \lesssim -0.5$ the
detectability of cosmic topology, for fixed $\Omega_m^{eff}$, is
much less likely than in the $\Lambda$CDM-dominated universe. We
note that the relevant features of this figure do not change
significantly for $\Omega_{b0}=0$, keeping $\Omega_m^{eff}=0.3$, of
course.

\begin{figure}[!hbt]
\centerline{\def\epsfsize#1#2{0.9#1}\epsffile{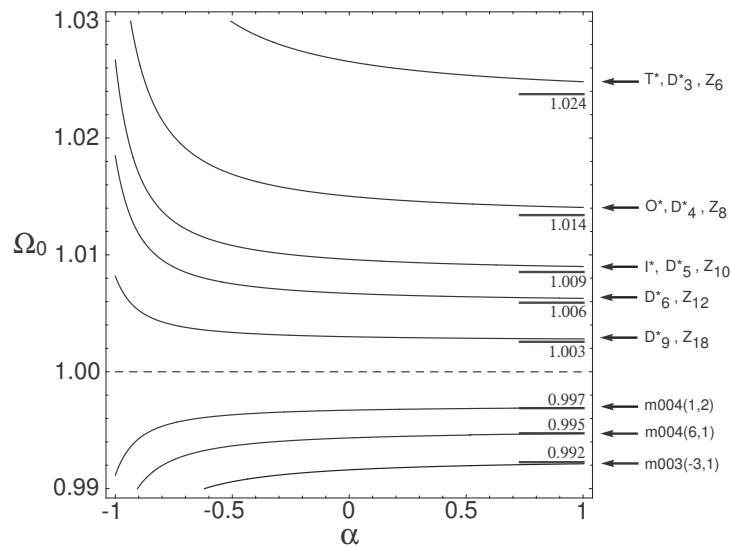}}
\caption{\label{OmegaVersusAlfa} The solution curves of
$\chi_{obs}(\Omega_0,\alpha)=r_{inj}$ for some spherical and small
hyperbolic manifolds of tables~\ref{SingleAction} and
\ref{10HW-Census}. An effective matter density $\Omega_m^{eff} =
0.3$, with $\Omega_{b0} = 0.04$, and a redshift $z_{obs}= 1089$ were
used.}
\end{figure}

To close this section we consider some specific manifolds of tables
~\ref{SingleAction} and~\ref{10HW-Census}, and focus our attention
on the parameter plane $\Omega_0$--$\,\alpha$. Taking again
$\Omega_m^{eff}=0.3$, with $\Omega_{b0}=0.04\,$, for $z_{obs}=1089$,
and $\Omega_0$ within the previously discussed bounds, we have
plotted the curves solution of~(\ref{ContCurves}) for some spherical
and hyperbolic manifolds, as shown in figure~\ref{OmegaVersusAlfa}.
From this figure we have that the greater the order of the group for
the cyclic and binary dihedral families the weaker is the
restriction on $\alpha$. So, for example, for $n>10$ and $m>5$ the
detectability of a given topology of these families would set no
restriction on the GCG parameter $\alpha$. An important point that
can be inferred from this figure is that if no sign is found of a
given topology in the $\Lambda$CDM framework this does not
necessarily mean that such topology is ruled out by observation. For
example, if no sign of either a binary tetrahedron $T^*$ or the
Weeks manifold m003(-3,1) is found in CMBR maps, it can be that
$\alpha \lesssim -0.4$ such that these topologies are unobservable
using pattern repetitions. Finally, if future observations such as
the Planck~\cite{Planck} mission tighten the range of $\Omega_0$
even closer to $1$, as for example $\Omega_0 = 1 \pm 0.003$, none of
the topologies considered in this figure would be detectable, for
any value of $\alpha$.

\section{Combination with Supernovae Data}
\label{combSNIa}
In the preceding section, we have investigated the constraints on
the matter content of the universe placed by the detection of a
nontrivial topology. Very often in cosmology the measurement of a
single observable does not allow to place strong limits on the
parameters of a model. It is therefore worthwhile to consider the
constraints that could arise from the detection of the topology
together with other sets of observational data. We shall consider in
this section the constraints on the GCG parameters which arise from
the detection of the cosmic topology combined with the limits
imposed by recent supernovae data (194 SNIa from
refs.~\cite{tonry03} and~\cite{Barris04}). We shall follow a similar
procedure as in section~\ref{DetectInver}, superimposing the
$\chi_{obs}=r_{inj}$ curves corresponding to some spherical and
hyperbolic manifolds to the regions in the parameter space allowed
by supernovae data.

\begin{figure}[!hbt]
\centerline{\def\epsfsize#1#2{0.9#1}\epsffile{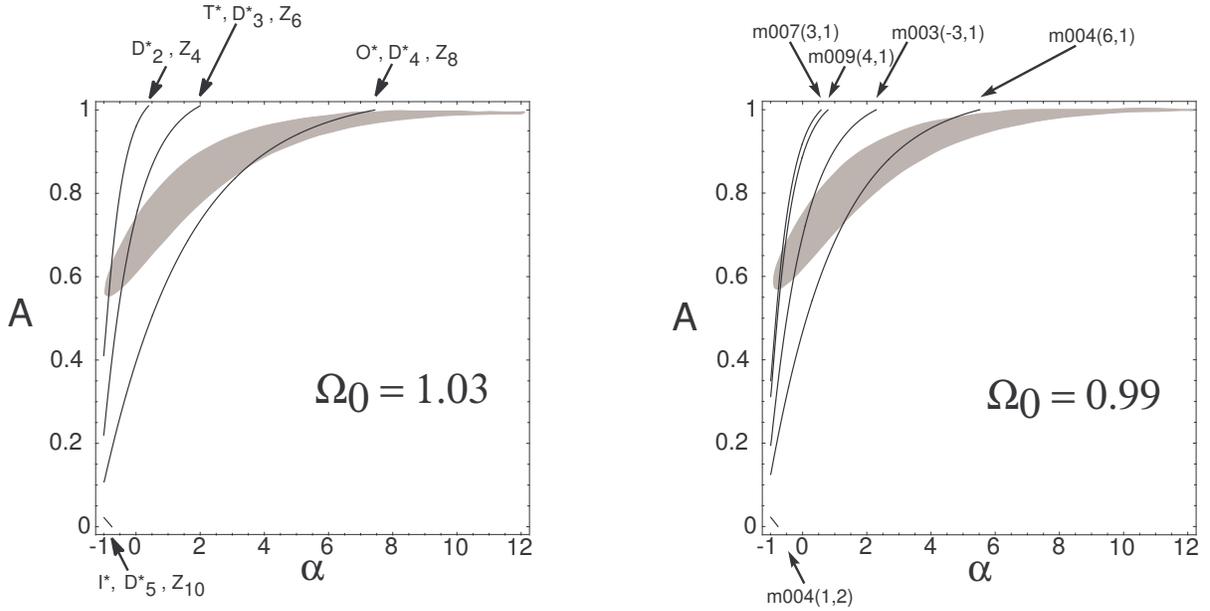}}
\caption{\label{AVersusAlfaSN} Superposition of the $95\%$
confidence regions from SNIa data (light gray) and the curves
$\chi_{obs}=r_{inj}$ for some spherical (left) and
hyperbolic (right) manifolds, and for a survey depth of
$z_{obs}= 1089$, as in figure~\ref{AVersusAlfa}.}
\end{figure}

In figure~\ref{AVersusAlfaSN} we display the $95\%$ ($2\sigma$)
confidence regions in the $A-\alpha$ parametric plane, for $\Omega_0
= 1.03$ (left) and $\Omega_0 = 0.99$ (right), from SNIa data. The
contours of constant confidence are nearly identical in both cases,
since supernovae data are not very sensitive to the total density.
We refer the readers to the appendix for a discussion on how these
contours are derived. The curves solution of $\chi_{obs}= r_{inj}$
are shown for some spherical (left) and hyperbolic (right)
manifolds. It is clear from the figure that the higher the value of
$r_{inj}$, the tighter are the combined constraints on both $A$ and
$\alpha$ (recall that only in models whose parameters are above
these curves the corresponding topology is detectable). For
instance, a detection of the dihedral $D^*_2$ (or $Z_4$) topology,
together with the SNIa constraints, would imply $ -1 \lesssim \alpha
\lesssim -0.5 $ and $ 0.55 \lesssim A \lesssim 0.65$ at $95\%$
confidence level. This constraint on $\alpha$ is of the same order
of what would be obtained with the Supernovae Acceleration
Probe~\cite{GCG1}. The detection of the $D^*_2$ topology would allow
to discard the $\Lambda$CDM model taking into account current
supernovae bounds. Incidentally, this makes clear that a given
topology that is undetectable in the $\Lambda$CDM model ($\alpha =
0$) may be detectable for other matter content models, as in the GCG
case~\cite{MMR1}. The detection of a tetrahedral $T^*$ (as well as
$D^*_3$ or $Z_6$) would set the constraints  $ -1 \lesssim \alpha
\lesssim 0 $ and $ 0.55 \lesssim A \lesssim 0.75$ at $95\%$
confidence level. The determination of the octahedral topology (as
well as $D^*_4$ or $Z_8$) would rule out only high values of
$\alpha$ (setting $\alpha \lesssim 7$). As expected from the
discussion of the previous section, for $\Omega_0=1.03$, the
Poincar\'e dodecahedral does not add limits to the supernovae
bounds, nor do the dihedral and cyclic families for $n>4$ and $m>8$.

Similarly, from the left panel of figure~\ref{AVersusAlfaSN}, the
detection of the higher $r_{inj}$ manifolds of
table~\ref{10HW-Census}, namely m$007(3,1)$, m$009(4,1)$, and
m$003(-3,1)$, would set rather strong constraints on $\alpha$. The
detection of m$004(6,1)$ topology gives rise to the bound $\alpha
\lesssim 5.5$, restricting therefore the limits set by supernovae
data on this parameter. As for the topology with the smallest
$r_{inj}$ of table~\ref{10HW-Census}, m$004(1,2)$, its detection
does not add new bounds to those which arise from supernovae.

Let us now investigate what happens when $\Omega_0$ is allowed to
vary within the bound $0.99<\Omega_0 < 1.03$ for some noteworthy
values of $\alpha$. We shall take the following values for $\alpha$.
First $\alpha  = -1/2$, which is a lower  limit set by the
combination of several observables assuming a flat universe.%
\footnote{This corresponds to the $2\sigma$ lower limit
derived in ref.~\cite{GCG2} and at $3\sigma$ in~\cite{ZhuGCG}.}
Second, $\alpha  = 0$, which corresponds to the $\Lambda$CDM case.
Third, $\alpha  = 1$, which is the standard Chaplygin gas.
Finally, we shall take $\alpha  = 2$, which is approximately the
best fit from the supernovae data for fixed $\Omega_0$ in the
interval $0.99<\Omega_0 < 1.03$.

\begin{figure}[!bht]
\centerline{\def\epsfsize#1#2{1.1#1}\epsffile{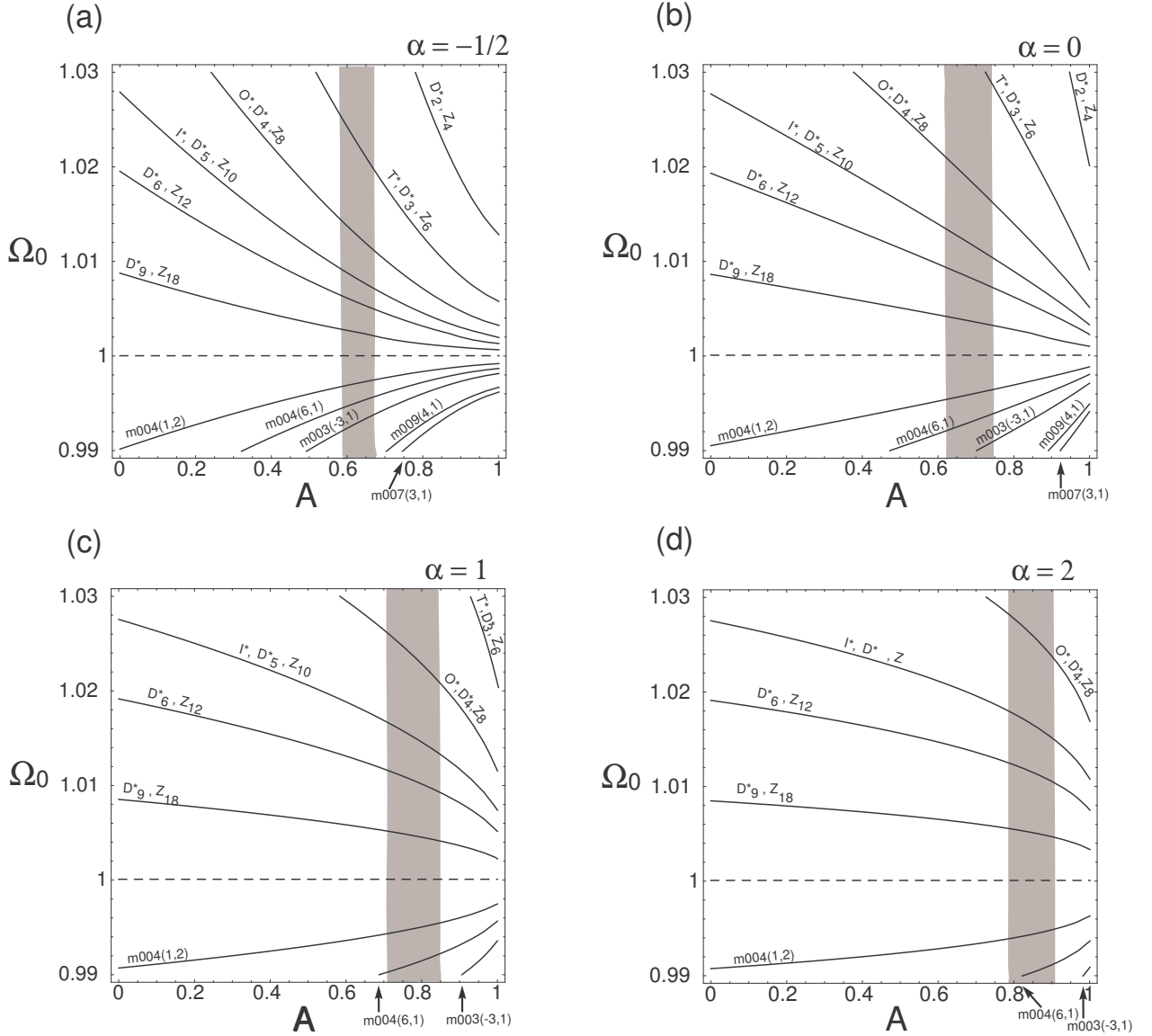}}
\caption{\label{OmegaVersusASN} Superposition of the $95\%$
confidence regions from SNIa data (light gray) and the curves
$\chi_{obs}=r_{inj}$ for several spherical and hyperbolic manifolds,
and for a survey depth of $z_{obs}= 1089$, as in figure
\ref{OmegaVersusA}.}
\end{figure}

In figure~\ref{OmegaVersusASN}, the $95\%$ confidence regions from
supernovae data for these values of $\alpha$ are shown superimposed
with the solution of $\chi_{obs}= r_{inj}$ for several manifolds. As
expected, the supernovae constraints on $A$ are almost independent
of $\Omega_0$ within the narrow interval of $\Omega_0$ that we are
considering. Therefore, the determination of a nontrivial topology
sets no further constraints on $A$ than those arising from the
supernovae data. However, knowing the topology, in combination with
the supernovae limits, allows to constrain the total density
$\Omega_0$. Indeed, for $\alpha = -1/2$, for example, if the
tetrahedral $T^*$ (or $D^*_3$, or $Z_6$) is found to be the cosmic
topology, this sets the constraint $\Omega_0 \gtrsim 1.02$. It
should be noticed that, according to SNIa data, a few topologies of
tables~\ref{SingleAction} and~\ref{10HW-Census} are already
unobservable at $2\sigma$ ($95\%$ confidence) for $\alpha=-1/2$ and
$0.99<\Omega_0<1.03$, such as $D^*_2$, $Z_4$, m$009(4,1)$, and
m$007(3,1)$. The $T^*$ topology is undetectable for $\alpha \gtrsim
0$, while the Poincar\'e dodecahedron $I^*$ is detectable for any
$\alpha$ in the considered range.

For a given spherical topology, the greater is $\alpha$ the higher
is the lower bound on $\Omega_0$ which arises from the detection of
the corresponding topology combined with SNIa data. Thus, e.g.,
while for $\alpha = -1/2$ the detection of the tetrahedral $T^*$
sets the bound $\Omega_0 \gtrsim 1.02$, for $\alpha= 0$ it gives
rise to the lower bound $\Omega_0 \gtrsim 1.029$.

As far as hyperbolic topologies are concerned, the greater is
$\alpha$ the smaller is the value for lower bound on $\Omega_0$
which arises from the combination of the topology detection with
SNIa data. Thus, for example, for the manifold m$004(6,1)$ and for
$\alpha = -1/2$ and $\alpha=1$ the lower bounds are $\Omega_0
\gtrsim 0.995$ and $\Omega_0 \gtrsim 0.990$, respectively. For
smaller $r_{inj}$ manifolds, however, such as $D^*_9$, $Z_{18}$, and
m$004(1,2)$, the change of the lower bound on $\Omega_0$ with
$\alpha$ is not very significative, as can be seen from the figures.
This allows to set constraints which are rather independent of
$\alpha$, for these manifolds. For instance if the topology of the
universe is found to be the m$004(1,2)$ manifold, then we must have
$\Omega_0 \lesssim 0.998$ at $95\%$ confidence level, for any value
of $\alpha \gtrsim -1/2$.

\section{Final Remarks}
\label{Conclusion}

Regardless of our present-day inability to predict the topology of
the universe, its detection and determination is ultimately expected
to be an observational problem. The search for topological signs of
a nontrivial topology of the universe has become particularly
topical. A finite universe is not only consistent with the main
pillars of the FLRW cosmological model, but offers a possible
explanation for several CMBR anisotopy features which are unexpected
in the context of an infinite flat $\Lambda$CDM
model~\cite{Poincare,CMB+NonGauss,Aurich1,Aurich2}. Although it is
not clear whether evidences of topological pattern
repetitions have been found in CMBR anisotropies maps%
~\cite{Cornishetal03,Roukema,Aurich1,Luminet05}, it is expected that
we should be able to detect the cosmic topology in the future, given
the wealth of increasingly accurate cosmological observations,
especially the recent results from the WMAP, and the development of
new methods and strategies for such a detection.

The detection of a nontrivial cosmic topology would certainly be a
major scientific discovery. Besides its importance \emph{per se}, we
have shown in this article that knowledge of the cosmic topology can
be used to set bounds on the matter content of the universe. To
concretely illustrate this fact, we have considered the GCG model
for dark matter and dark energy unification and determined the
constraints on its parameters that would be obtained from the
detection of a number of topologies. Specifically, we have
considered the seven smaller volume hyperbolic manifolds and the
globally homogeneous spherical manifolds.

We have shown that for $\Omega_0$ within the current observational
bounds, the determination of cosmic topology generally sets upper
limits on $\alpha$ and lower limits on $A$ (see, for example,
figures~\ref{OmegaVersusA} and~\ref{AVersusAlfa}). Furthermore, the
constraints on $\alpha$ improve as the allowed interval of
$\Omega_0$ narrows, as expected from the forthcoming experiments
(see, e.g., figure~\ref{OmegaVersusAlfa}). Also, the higher the
injectivity radius $r_{inj}$, the stronger are the constraints from
the detection of the corresponding manifolds through pattern
repetitions.

We have also shown that the cosmic topology detection sets
additional constraints on the matter content parameters to the
bounds that arise from the SNIa data analysis, by using the GCG
parameters as a concrete example. To this end we have determined the
confidence regions in the $A$~--~$\alpha$ parameter plane from 194
SNIa for fixed $\Omega_0$ at the extreme values of the current
observational bounds, obtained the $95\%$ limits on the
$\Omega_0$--$A$ plane for some physically motivated values of
$\alpha$, and combined these limits with the bounds from cosmic
topology detection for several spherical and hyperbolic topologies.
As a result we found that this combination allows to set stringent
constraints on the GCG parameters. Thus, for example, the detection
of $D^*_2$ implies $-1 \lesssim \alpha \lesssim -0.5 $ and $0.55
\lesssim  A \lesssim 0.65$ at $95\%$ confidence level, for
$\Omega_0=1.03$. These constraints are stronger than those obtained
from any current data based on the background
dynamics~\cite{ZhuGCG}. Clearly, if limits from other sets of
observational data are combined with constraints from the
determination of the topology, the latter would greatly improve the
former bounds.

We found that, besides the GCG parameters, the detection of the
topology provides constraints on the total density $\Omega_0$.
First, the detection of a spherical (hyperbolic) topology obviously
implies $\Omega_0> 1$ ($\Omega_0 < 1$). Second, for each manifold
the topology detection implies a lower (upper) bound for spherical
(hyperbolic) $\Omega_0$. 
Finally, the combination of cosmic topology detection with
supernovae data further constrains these bounds on $\Omega_0$, by
restricting the allowed interval of $A$ for different values of
$\alpha$.

The determination of a possible nontrivial cosmic topology is a
problem that can be addressed with the current observational data.
Several searches have been conducted and more will be carried on in
the future. The fact that the determination of the topology allows
to constrain the matter-energy content of the universe provides an
extra motivation for this search.

\section*{Acknowledgements}
We thank MCT and CNPq for the grants under which this work was
carried out. We also thank A.F.F. Teixeira for the reading of the
manuscript and indication of relevant misprints and omissions. M.M.
acknowledges L.S. Werneck for support with the supernovae analysis,
I. Waga for useful suggestions, and FAPERJ for financial support.

\appendix*

\section{Constraints on the GCG from recent SNIa data}
\label{SNIaTop}

Empirical studies show that SNIa can be used as standard candles
after light curve calibration (see, e.g.,
refs.~\cite{SNreview,recentSNIa}). Therefore, they offer a direct
probe of~the luminosity distance-redshift relation, which can be
used to constrain theoretical models.

In section~\ref{combSNIa} we displayed confidence levels on the
parameters $A$ and $\alpha$ of the GCG model obtained from SNIa
data. To produce such contours, a sample of 253 SNIa was compiled
from references~\cite{tonry03} and~\cite{Barris04} which provide
tables with the redshift $z$, $\log D_{L}$, and its variance
$\sigma_{\log D_{L}}^{2}$ for each supernovae. Here $D_{L}$ is the
luminosity distance times the Hubble constant, which is given by
\begin{equation}
D_{L}=d_{L}H_{0}=\frac{1+z}{\sqrt{|1-\,\Omega_{0}|}}\,S_k
\left(\sqrt{|1-\,\Omega_{0}|}\,H_{0}\int_{0}^{z}\frac{dz^{\prime}}%
{H(z^{\prime})}\right)
=\frac{1+z}{\sqrt{|1-\,\Omega_{0}|}}\,S_k(\chi)\;,
\label{DL}%
\end{equation}
where $S_k(x) = \sin(x)$,  $\sinh(x)$,  or $x$, for $\Omega_0$
greater, smaller or equal to unity, respectively. The Hubble
function $H(z)$ is given by eq.~(\ref{HubbleFun}) and $\chi$ is
given by~(\ref{ChiObsGCG1}). As mentioned in the text, we fix
$\Omega_{b0}=0.04$, in agreement with the observed abundances of
light elements and measurements of the Hubble constant.

Following~\cite{tonry03} and~\cite{Barris04} we discard local
supernovae with $z<0.01$, because the peculiar motion contribution
to $z$ is too high, and those with high host extinction,
$A_{V}>0.5$, which could cause a strong bias in the determination of
$D_{L}$. After these cuts, we end up with a sample of $194$ SNIa
extending up to $z=1.75$. In computing the Chi-squared, we take into
account the scatter in $z$ caused by peculiar velocities. Assuming a
velocity dispersion of $\sigma_{v}=500Km/s$, we propagate the
uncertainty in $z$ into the luminosity distance, adding the result
in quadrature with the observational uncertainty in $D_{L}$.
Therefore, the Chi-squared is given by:
\[
\chi^{2}=\sum_{i=1}^{194}\frac{\left[  \log\left(  D_{L}^{Obs}\left(
z_{i}\right)\right)-\log\left(D_{L}^{Th}\left(z_{i}\right)\right)
\right]^{2}}{\sigma_{\log\left(  D_{L}\left(z_{i}\right)\right)}
^{2}+\left(\left.\frac{\partial\log D_{L}^{Th}}{\partial
z}\right\vert _{z_{i}}\sigma_{z}\right)^{2}}\,,
\]
where the theoretical prediction is computed from (\ref{DL})
together with (\ref{ChiObsGCG1}) and the observational values are
given in the tables of~\cite{Barris04} and~\cite{tonry03}.

To generate the contours displayed in
section~\ref{combSNIa}, we assume that the 
$95\%$ confidence levels are well approximated by the same value of
$\Delta \chi^2$ 
as for a two-dimensional normal distribution. That is, we obtain the
best fit (minimum of $\chi^2$) and plot the contour levels of
$\Delta \chi^2=
6.17$.

To produce the confidence regions of figure
\ref{AVersusAlfaSN}, we fixed $\Omega_0=1.03$ and $\Omega_0=0.99$
for the spherical and hyperbolic universes. The parameters $A$ and
$\alpha$ are allowed to vary, with no priors. On the other hand, for
the graphs of figure~\ref{OmegaVersusASN} we fix $\alpha$ at several
physically motivated values and leave $\Omega_0$ and $A$ totaly
free. The results are shown only for a narrow interval of $\Omega_0$
allowed by a combination of several observables in the context of the
$\Lambda$CDM model, namely $0.99<\Omega_0<1.03$.

As far as we know, this is the first time that such analysis is done
for the GCG model. By fixing $\Omega_0$ motivated by other non SNIa
data, one is able to obtain stronger limits on $A$ and $\alpha$ then
when $\Omega_0$ is left completely free. As expected, these limits
are almost insensitive to $\Omega_0$ in the range
$0.99<\Omega_0<1.03$, i.e., they are robust for nearly flat
geometries. For fixed $\alpha$, the constraints on $A$ are almost
independent of $\Omega_0$ in the range above, as can be seen in
figure~\ref{OmegaVersusASN} [section~\ref{combSNIa}]. It is clear
that SNIa data alone cannot place significative constraints on
$\Omega_0$ within that narrow range (see, e.g.,
ref.~\cite{FabrisSNIa}).

\end{document}